\documentclass[showpacs,showkeys,preprintnumbers,amsmath,amssymb]{revtex4}
\usepackage[dvips]{graphicx}
\newcommand{\be}{\begin{equation}}
\newcommand{\ee}{\end{equation}}
\newcommand{\bea}{\begin{eqnarray}}
\newcommand{\eea}{\end{eqnarray}}
\newcommand{\bm}{\begin{mathletters}}
\newcommand{\eml}{\end{mathletters}}
\newcommand{\nn}{\nonumber\\}

\begin{document}

\title{SYSTEMATICS OF PROTON EMISSION}

\author{D.S. Delion}
\affiliation{
National Institute of Physics and Nuclear Engineering, \\
POB MG-6, Bucharest-M\u agurele, Romania}
\author{R.J. Liotta and R. Wyss}
\affiliation{
KTH, Alba Nova University Center, SE-10691 Stockholm, Sweden}
\date{\today}

\begin{abstract}

A very simple formula is presented that relates the logarithm of the
half-life, corrected by the centrifugal barrier, with the Coulomb
parameter in proton decay processes.
The corresponding experimental data lie on two straight lines
which appear as a result of a sudden change in the nuclear shape marking
two regions of deformation independently of the angular momentum
of the outgoing proton.
This feature provides a powerful tool to assign experimentally
quantum numbers in proton emitters.
\end{abstract}



\maketitle

%\newpage
\setcounter{equation}{0}
\renewcommand{\theequation}{\arabic{equation}}

Nearly a century ago one of the most challenging problems in physics was
the understanding of the Geiger-Nuttall rule, which says that the logarithm
of the half-life in alpha decay is inverse proportional to the energy of the
outgoing alpha particle, i. e. its Q-value. It can be asserted that the
probabilistic interpretation of quantum mechanics started by the
explanation given by Gamow to that rule as a consequence of the quantum
penetration of the alpha particle through the Coulomb
barrier \cite{Gam28}. Nowadays the challenge in nuclear physics
is related to rare nuclei, i. e. nuclei lying very far from the stability
line which decay rapidly by particle emission (neutron and proton drip
lines) \cite{Woo97}. The exploration of the drip-lines is one of the ambitions
of the projected radioactive nuclear beam facilities as e. g. the Rare Isotope
Accelerator (RIA) \cite{arg}.

By plotting the logarithm of the half-lives as a function of $Q_p$, i. e.
the kinetic energy of the emitted proton, one does not obtain a clear
graphical pattern of the experimental data, as for the Geiger-Nuttall rule.
As seen in Figure 1.a \cite{Son02}, such a plot does not
reveal any obvious trend.
The reason of this disorder is that not only $Q_p$ but also the height
and width of the Coulomb and centrifugal barriers, in which the proton
is trapped before decaying, determine the decay width.
The role of the centrifugal barrier in proton emission is more important
than in $\alpha$-decay due to the smaller proton reduced mass
and also because in most cases the proton carries a non-vanishing angular momentum.
The half-life corresponding to the wave with given quantum numbers $(l,j)$
can be written as \cite{Fro57}
\bea
\label{T}
T_{1/2}&=&\frac{ln 2}{v}\left|\frac{H^{(+)}_l(\chi,\rho)}
{s_{lj}(R,\beta)}\right|^2~,
\eea
where $H^{(+)}_l$ is the Coulomb-Hankel spherical wave,
$\chi=2(Z-1)e^2/(\hbar v)$ is the Coulomb parameter
(which determines the Coulomb barrier),
$Z$ is the charge number of the mother nucleus,
$v=\hbar k/\mu=\sqrt{2 Q_p/\mu}$ is the velocity of the outgoing proton,
$\mu$ is the reduced mass of the proton-daughter system, $\rho=kR$ and
the spectroscopic function is given by \cite{Del05}
\bea
\label{s}
s_{lj}(R,\beta)&=&\sum_{l'j'}
K_{lj;l'j'}(R,\beta)s^{(0)}_{l'j'}f_{l'j'}(R,\beta)~,
\eea
where $s^{(0)}_{lj}$ is the so-called spectroscopic factor
(the particle amplitude $u_{lj}$ within the super fluid model) and
$f_{lj}$ are the components of the internal wave function
(the standard Nilsson function within the adiabatic approach).
We mention here that $|s^{(0)}_{lj}|^2\approx 0.5$. Moreover for spherical
emitters it was evidenced a clear dependence of the spectroscopic factor
upon the shell occupancy \cite{spf}.
The asymptotic propagator matrix is given by
\bea
\label{K}
K_{lj;l'j'}(R,\beta)&\equiv&
H^{(+)}_l(\chi,\rho)\left[{\cal H}^{(+)}(R,\beta)\right]^{-1}_{lj;l'j'}
=\delta_{ll'}\delta_{jj'}+\Delta K_{lj;l'j'}(R,\beta)~.
\eea
Here ${\cal H}^{(+)}_{lj;l'j'}(R,\beta)$ is the matrix of solutions with
an outgoing boundary behavior, i.e.
${\cal H}^{(+)}_{lj;l'j'}(R,\beta)\rightarrow_{R\rightarrow\infty}
\delta_{ll'}\delta_{jj'}H^{(+)}_l(\chi,\rho)$.
%It is fully determined by the concrete form of the interaction.
Thus, the relation (\ref{T}) has the same form as in the
spherical case, but for deformed systems the spectroscopic
function is a superposition of the different channel components $f_{lj}$.
%Our coupled channels calculations showed that, even for the most
%deformed emitter $^{131}$Eu with $\beta=0.33$, the nondiagonal contribution
%is relative small, i.e. $max(\Delta K)< 0.1$, for distances at which the
%nuclear part of the interaction has vanishing values.

The ratio $H^{(+)}_l/s_{lj}$, entering Eq. (\ref{T}) (which is just
the inverse of the scattering amplitude in the $(l,j)$ channel) does
not depend upon the radius and this is an important test of accuracy
in any coupled channels scheme.
A good approximation of the function $H^{(+)}_l$
for physical situations is given by the WKB value, i. e. \cite{Fro57}
\bea
\label{H}
H^{(+)}_l(\chi,\rho)\approx G_l(\chi,\rho)\approx
C_l(\chi,\rho)(ctg~\alpha)^{1/2}
exp[\chi(\alpha-sin~\alpha~cos~\alpha)]~,
\eea
where $G_l$ is the irregular Coulomb function. The influence of the
centrifugal barrier is fully contained in the function $C_l$, which is
given by
\bea
\label{cl}
C_{l}(\chi,\rho) = exp\left[\frac{l(l+1)}{\chi}tg~\alpha\right]~,
~~~cos^2~\alpha =\frac{Q_p}{V_c(R)}=\frac{\rho}{\chi}~,
\eea
$V_c(R)$ is the Coulomb potential at distance $R$.
We will choose this distance as the matching radius,
for which we will adopt the standard form, i. e.
$R=1.2(A_d^{1/3}+A_p^{1/3})$, where $A_d$ is the mass number of the
daughter nucleus and $A_p=1$.
The dependence of the centrifugal factor upon the distance, entering
through $\alpha$, is very weak around that value of R.
Defining a reduced half-life as
\bea
\label{tred}
T_{red}=\frac{T_{1/2}}{C_l^2}=\frac{F(\chi,\rho)}{|s_{lj}(R,\beta)|^2}~,
\eea
where
\bea
\label{funcf}
F(\chi,\rho)=\frac{ln 2}{v}ctg~\alpha~
exp[2\chi(\alpha-sin~\alpha~cos~\alpha)]~,
\eea
one sees that $T_{red}$ should not depend upon the angular momentum $l$
if plotted against the dimensionless Coulomb parameter $\chi$.
Moreover, $log_{10}~T_{red}$ 
%which, according to Eq. (\ref{funcf}),
%is proportional to $2\chi(\alpha-sin~\alpha~cos~\alpha)$,
is a linear function of $\chi$ independently of the value of $l$,
provided the velocity $v$, the function $s_{lj}$ and the parameter $\alpha$,
which depend upon the Q-value, has a smooth behavior in this logarithmic scale.
Notice that the value of $T_{red}$ is model
independent since the only theoretical quantity entering in its definition
is the function $C_l^2$, which is a by-product of the barrier penetration
process. But it is important to stress that this entire analysis is
based upon the assumption that the proper
value of $l$, that is the one that determines the experimental
half-life $T_{1/2}$, is used. Otherwise, and since $T_{1/2}$
is strongly $l$-dependent, that straight line pattern would be completely spoiled.

To check the rather straightforward conclusions reached above we evaluated
$log_{10}~T_{red}$ in cases where experimental data are available,
as shown in Table 1. We considered proton emitters with $Z>50$
and angular momentum as given in Ref. \cite{Son02}.
The orbital and total angular momenta of the proton in the mother nucleus
labeling a given partial wave will be denoted by $(l_m,j_m)$.
The corresponding outgoing values (at large distances) will be $(l,j)$.
Angular momentum conservation requires that
$\vec J_m=\vec J_d +\vec j$, where $J_m$ ($ J_d$) is the angular
momentum of the mother (daughter) nucleus.
In the case of odd decaying nuclei to the ground state with
$J_d=0$, it follows that $J_m=j_m$. If, in addition, the nucleus is spherical then
$j_m=j$ and $l_m=l$, where $l_m$ is the orbital angular momentum of the
single quasi-proton state in the mother nucleus.
The value of $j_m$ in Table 1 is the angular
momentum of the proton outside the nucleus. It is
often extracted from theoretical predictions.

With the values of $l$, $\chi$ and $log_{10}~T_{red}$ of
Table 1 we produced the plot shown in Figure 1.b.
Amazingly enough, the points lie all approximately along two straight lines.
The nuclei on the upper line correspond
to the first twelve cases in Table 1, i. e. those with $Z<68$.
It is important to point out that the angular momenta in all cases of Figure 1.b 
are as reported in the literature, except   
the nucleus $^{109}$I for which we changed the angular
momentum $l=0$ of Ref. \cite{Son02} to $l=2$. We did this 
because we performed
a coupled channels calculation showing a larger component
with l=2 (0.37) in comparison with l=0 (0.02) for the 1/2+ deformed state
at the Fermi level. 
Note that this assignment is also in agreement with the systematics
in heavier isotopes, like e.g. in $^{111}$I\cite{Pau00}.

Using a fitting procedure we found that the experimental
half-lives can be reproduced by the formula
\bea
\label{lines}
log_{10}~T_{red}^{(k)}&=&a_k(\chi-20)+b_k~,
\nn
a_1&=&1.31~,~~~b_1=-2.44~,~~~Z<68
\nn
a_2&=&1.25~,~~~b_2=-4.71~,~~~Z>68~.
\eea
where $k=1$ corresponds to the upper line in Figure 1.b.
The standard errors are $\sigma_1=0.26$ and $\sigma_2=0.23$,
corresponding to a mean factor less than two.

These two straight lines may have been induced
either by an abrupt change in the Q-values or in
the structure of the different emitters, or by both.
The Q-value dependence affects only the function $F(\chi,\rho)$.
We therefore plotted this function against $\chi$ in Figure 2.
One sees in this Figure the same pattern as in Figure 1.b,
namely there appear two lines, but now the upper line includes
only six emitters with $Z<68$.
In fact the deviations from the two lines in Figure 2
are much larger here than in Figure 1.b, corresponding
to an error of about one order of magnitude.
This forces us to conclude that, although the pattern shown in Figure 2
corresponding to $s_{lj}(\beta)=const$ is similar to the dependence between
$\rho$ and $\chi$, this effect can not be the only reason behind
the two straight lines of Figure 1.b. The only other
source that may contribute to that alignment is an abrupt change in the
nuclear structure of the emitters.
We therefore correlated the deformation
values calculated in Ref. \cite{Mol95} with the determined half-lives.
In Figure 3.a we show the deformation parameter $\beta$ as a function of Z.
At the proton drip line between Z=67 and Z=69 occurs a pronounced change,
from a large prolate shape with $\beta\approx$ 0.3 to an oblate shape
with $\beta\approx$ -0.2. These shapes are substantiated by measurements
of moments of inertia.
In order to analyze the dependence of the half-life upon the internal
structure of the nucleus
we plot in Figure 3.b the quantity $log_{10}s_{lj}^{-2}$ as a function
of the charge number Z. It is worthwhile to emphasize once again
that this is a model independent function.

One remarks a striking correlation between Figures 3.a and 3.b.
After the jump occurring at Z=68, where an abrupt shape change occurs,
the deformation of the nuclei lying on the lower line smoothly increases. 
One can then assert that these lines reflect two regions of nuclei
separated by a sharp transition between the prolate and oblate regimes.

There are two points in Figure 1.b that deviate conspicuously from the
upper straight line. They correspond to $^{140}$Ho (l=3) and
$^{141}$Ho$^*$ (l=0). In Figure 3.a one sees that these nuclei
are situated at the border between the two regions of deformation.

In conclusion we have presented in this paper a simple formula for
proton decay (Eq. (\ref{lines})) similar to
the Geiger-Nuttall rule. This formula enables the precise assignment 
of spin and parity for proton decaying states.
The only quantities that are needed are the half-life of the mother
nucleus and the proton Q-value. As a function of these quantities, corrected
by the centrifugal barrier (Eq. (\ref{tred})), the experimental data
of proton emitters with $Z>50$ lie along two straight lines.
Taking into account that $^{140}$Ho $^{141}$Ho$^*$ 
are the only emitters
that deviates from the systematics, one concludes the remarkable fact:
proton emitters with $Z>50$ follow simple systematics, 
which are obviously correlated with two regions of quadrupole deformation,
as can be seen by plotting $log_{10}s^{-2}$ versus $\beta$.

\acknowledgments
This work has been supported by the G\"oran Gustafsson Foundation.

%\newpage

\newpage
\vskip0.5cm
\begin{center}
{\bf Table 1}
\vskip0.5cm
{\it Data used in the calculation. The spin of
the proton moving inside the mother nucleus ($j_m$),
the proton Q-value ($Q_p$) and the emitter half-life
($T_{exp}$) are from Refs. \cite{Son02,Woo04,Rob05}.
The quadrupole deformation parameters $\beta$
are from Ref. \cite{Mol95} and the quantities $l$, $\chi$, $\alpha$,
$T_{red}$ and $s_{lj}$ are
as explained in the text. All half-lives are in seconds.
Stars in the emitters indicate excited states.}
\vskip0.5cm
\begin{tabular}{|c|c|c|c|c|c|c|c|c|c|c|}
\hline
~~No~~ & Emitter & ~~$l$~~ & ~~$j_m^\pi$~~ & $Q_p$ &
$\chi$ & $\beta$ & $\alpha$ & $log_{10}~T_{exp}$ &
$log_{10}~T_{red}$ & $log_{10}~s_{lj}^{-2}$ \cr
& & & & (keV) & & & (rad) & & & \cr
\hline
~1&$^{105}_{~51}$Sb$^{~}$ & 2&~5/2$^{+}$& ~491(15)& 22.457& ~0.081& 1.353& ~2.049& ~1.000 & 1.525 \cr
~2&$^{109}_{~53}$I~$^{~}$ & 2&~3/2$^{+}$& ~829(~3)& 17.977& ~0.160& 1.290& -3.987& -4.994 & 1.810 \cr
~3&$^{112}_{~55}$Cs$^{~}$ & 2&~3/2$^{+}$& ~824(~7)& 18.728& ~0.208& 1.296& -3.301& -4.287 & 1.701 \cr
~4&$^{113}_{~55}$Cs$^{~}$ & 2&~3/2$^{+}$& ~978(~3)& 17.191& ~0.207& 1.270& -4.777& -5.747 & 1.213 \cr
~5&$^{117}_{~57}$La$^{~}$ & 2&~3/2$^{+}$& ~823(~5)& 19.437& ~0.290& 7.052& -1.602& -2.565 & 2.680 \cr
~6&$^{121}_{~59}$Pr$^{~}$ & 2&~3/2$^{+}$& ~900(10)& 19.253& ~0.318& 7.119& -2.000& -2.939 & 2.756 \cr
~7&$^{130}_{~63}$Eu$^{~}$ & 2& 3/2$^{+}$& 1028(15)& 19.263& ~0.331& 7.263& -3.046& -3.941 & 2.146 \cr
~8&$^{131}_{~63}$Eu$^{~}$ & 2&~3/2$^{+}$& ~951(~7)& 20.028& ~0.331& 1.289& -1.575& -2.473 & 2.592 \cr
~9&$^{135}_{~65}$Tb$^{~}$ & 3& 7/2$^{-}$& 1188(~7)& 18.499& ~0.325& 7.341& -3.027& -4.770 & 2.548 \cr
10&$^{140}_{~67}$Ho$^{~}$ & 3& 7/2$^{-}$& 1106(10)& 19.775& ~0.297& 7.416& -2.222& -3.937 & 1.864 \cr
11&$^{141}_{~67}$Ho$^{~}$ & 3&~7/2$^{-}$& 1190(~8)& 19.064& ~0.286& 1.261& -2.387& -4.094 & 2.672 \cr
12&$^{141}_{~67}$Ho$^{*}$ & 0&~1/2$^{+}$& 1256(~8)& 18.557& ~0.286& 1.252& -5.180& -5.180 & 2.268 \cr
13&$^{145}_{~69}$Tm$^{~}$ & 5&11/2$^{-}$& 1753(10)& 16.185& -0.199& 7.490& -5.409& -9.500 & 1.307 \cr
14&$^{146}_{~69}$Tm$^{~}$ & 5&11/2$^{-}$& 1210(~5)& 19.481& -0.199& 1.261& -1.276& -5.460 & 0.943 \cr
15&$^{146}_{~69}$Tm$^{*}$ & 5&11/2$^{-}$& 1148(~5)& 20.001& -0.199& 1.270& -0.456& -4.651 & 1.053 \cr
16&$^{147}_{~69}$Tm$^{~}$ & 5&11/2$^{-}$& 1071(~3)& 20.708& -0.190& 1.280& ~0.591& -3.613 & 1.148 \cr
17&$^{147}_{~69}$Tm$^{*}$ & 2&~3/2$^{+}$& 1139(~5)& 20.080& -0.190& 1.271& -3.444& -4.282 & 1.324 \cr
18&$^{150}_{~71}$Lu$^{~}$ & 5&11/2$^{-}$& 1283(~4)& 19.477& -0.164& 1.255& -1.180& -5.280 & 1.314 \cr
19&$^{150}_{~71}$Lu$^{*}$ & 2&~3/2$^{+}$& 1317(15)& 19.224& -0.164& 1.251& -4.523& -5.342 & 1.592 \cr
20&$^{151}_{~71}$Lu$^{~}$ & 5&11/2$^{-}$& 1255(~3)& 19.694& -0.156& 1.259& -0.896& -4.996 & 1.316 \cr
21&$^{151}_{~71}$Lu$^{*}$ & 2&~3/2$^{+}$& 1332(10)& 19.116& -0.156& 1.249& -4.796& -5.613 & 1.475 \cr
22&$^{155}_{~73}$Ta$^{~}$ & 5&11/2$^{-}$& 1791(10)& 16.958& ~0.008& 1.199& -4.921& -8.864 & 1.293 \cr
23&$^{156}_{~73}$Ta$^{~}$ & 2&~3/2$^{+}$& 1028(~5)& 22.384& -0.050& 1.292& -0.620& -1.433 & 1.448 \cr
24&$^{156}_{~73}$Ta$^{*}$ & 5&11/2$^{-}$& 1130(~8)& 21.350& -0.050& 1.278& ~0.949& -3.099 & 1.177 \cr
25&$^{157}_{~73}$Ta$^{~}$ & 0&~1/2$^{+}$& ~947(~7)& 23.322& ~0.045& 1.303& -0.523& -0.523 & 1.099 \cr
26&$^{160}_{~75}$Re$^{~}$ & 2&~3/2$^{+}$& 1284(~6)& 20.587& ~0.080& 1.261& -3.046& -3.837 & 1.649 \cr
27&$^{161}_{~75}$Re$^{~}$ & 0&~1/2$^{+}$& 1214(~6)& 21.172& ~0.080& 1.270& -3.432& -3.432 & 1.275 \cr
28&$^{161}_{~75}$Re$^{*}$ & 5&11/2$^{-}$& 1338(~6)& 20.167& ~0.080& 1.254& -0.488& -4.433 & 1.626 \cr
29&$^{164}_{~77}$Ir$^{~}$ & 5&11/2$^{-}$& 1844(~9)& 17.644& ~0.089& 1.201& -3.959& -7.768 & 1.838 \cr
30&$^{165}_{~77}$Ir$^{~}$ & 5&11/2$^{-}$& 1733(~7)& 18.201& ~0.099& 1.212& -3.456& -7.279 & 1.594 \cr
31&$^{166}_{~77}$Ir$^{*}$ & 5&11/2$^{-}$& 1340(~8)& 20.699& ~0.107& 1.257& -0.076& -3.955 & 1.578 \cr
32&$^{167}_{~77}$Ir$^{~}$ & 0&~1/2$^{+}$& 1086(~6)& 22.993& ~0.116& 1.289& -0.959& -0.959 & 1.492 \cr
33&$^{167}_{~77}$Ir$^{*}$ & 5&11/2$^{-}$& 1261(~7)& 21.338& ~0.116& 1.266& ~0.875& -3.012 & 1.670 \cr
34&$^{171}_{~79}$Au$^{~}$ & 0&~1/2$^{+}$& 1469(17)& 20.291& -0.105& 1.245& -4.770& -4.770 & 1.498 \cr
35&$^{171}_{~79}$Au$^{*}$ & 5&11/2$^{-}$& 1718(~6)& 18.763& -0.105& 1.217& -2.654& -6.414 & 1.901 \cr
36&$^{177}_{~81}$Tl$^{~}$ & 0&~1/2$^{+}$& 1180(20)& 23.223& -0.050& 1.282& -1.174& -1.174 & 1.342 \cr
37&$^{177}_{~81}$Tl$^{*}$ & 5&11/2$^{-}$& 1986(10)& 17.901& -0.053& 1.192& -3.347& -7.006 & 2.648 \cr
38&$^{185}_{~83}$Bi$^{~}$ & 0&~1/2$^{+}$& 1624(16)& 20.293& -0.052& 1.232& -4.229& -4.229 & 2.436 \cr
\hline
\end{tabular}
\end{center}

\newpage

\begin{figure}[p]
\begin{center}
\includegraphics[]{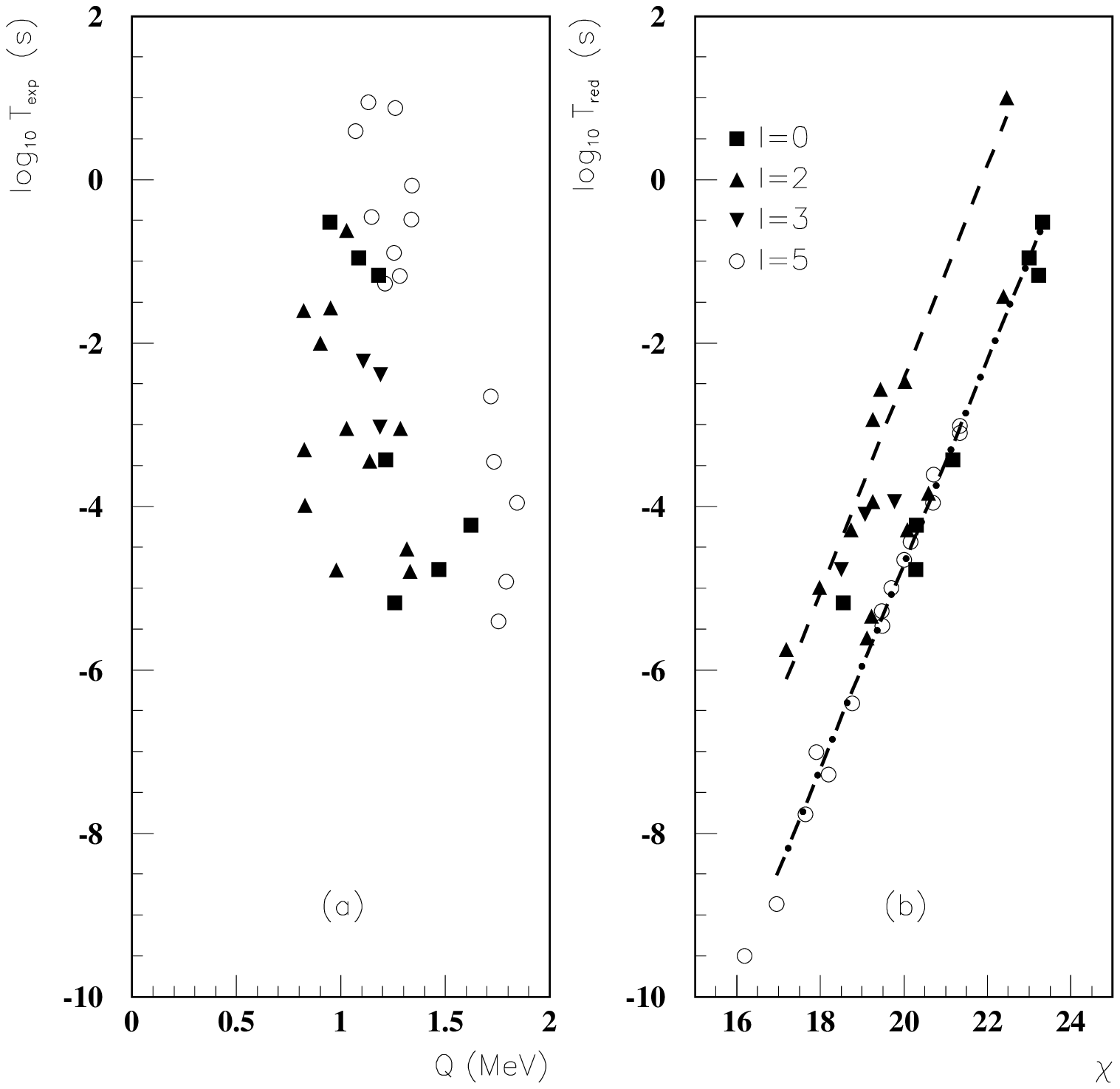}
\vspace{1cm} \caption
{
(a) Logarithm of the experimental half-lives corresponding to proton decay
as a function of the Q-value. The data are taken from Ref. \cite{Son02}.\\
(b) Values of $log_{10}~T_{red}$, Eq. (\ref{tred}),
as a function of the Coulomb parameter $\chi$,
The numbers labelling the different symbols correspond to
the $l$-values of the outgoing proton.
The two lines are computed according to Eq. (\ref{lines}).
}
\label{fig1}
\end{center}
\end{figure}

\newpage
\begin{figure}[p]
\begin{center}
\includegraphics[]{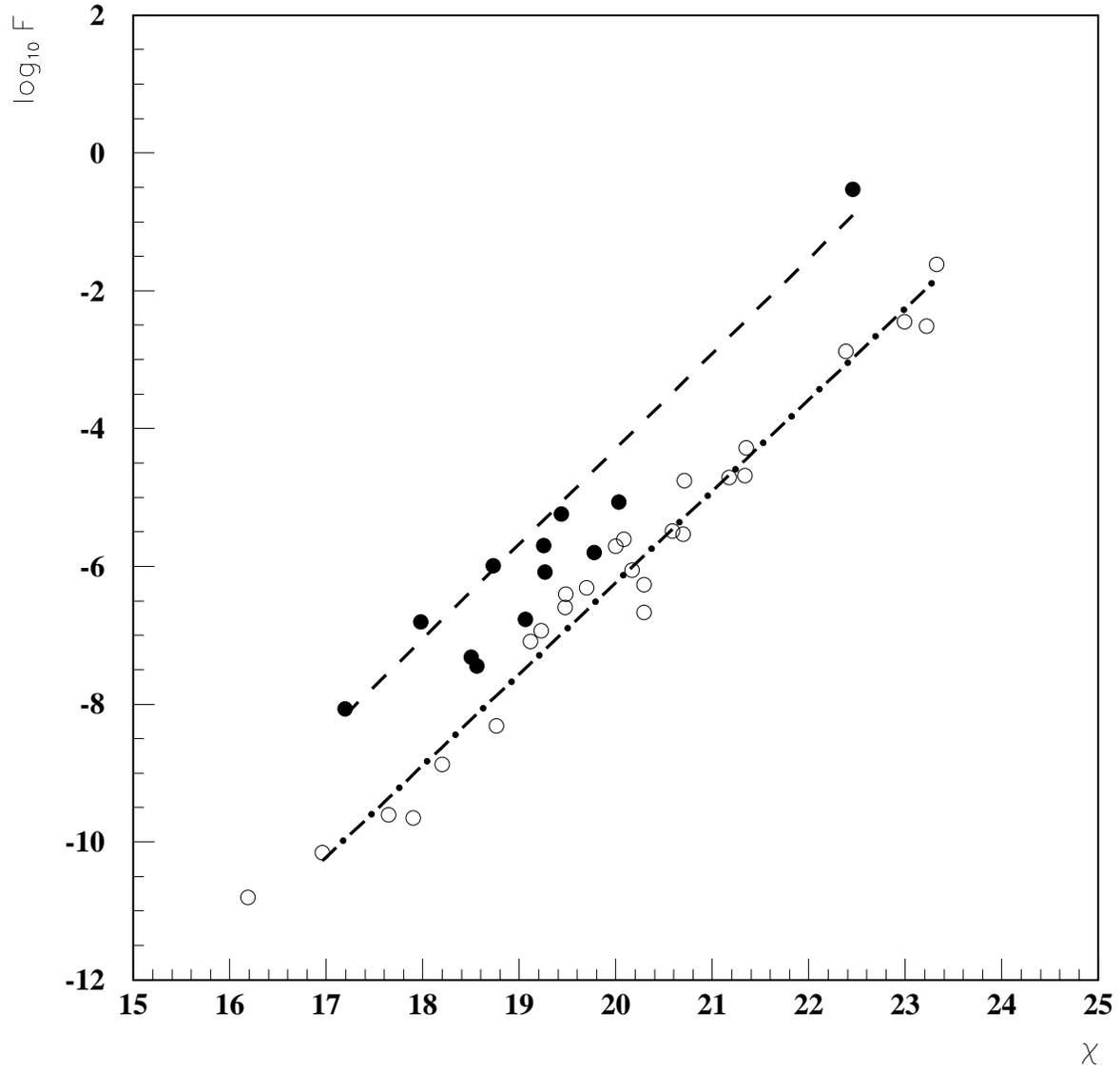}
\vspace{1cm} \caption
{The logaritm of $F(\chi,\rho)$ (s/fm), Eq. (\ref{funcf}),
as a function of $\chi$. The two straight lines are average values
to guide the eye. Filled symbols correspond to the first twelve cases
of Table 1.
}
\label{fig2}
\end{center}
\end{figure}

\newpage
\begin{figure}[p]
\begin{center}
\includegraphics[]{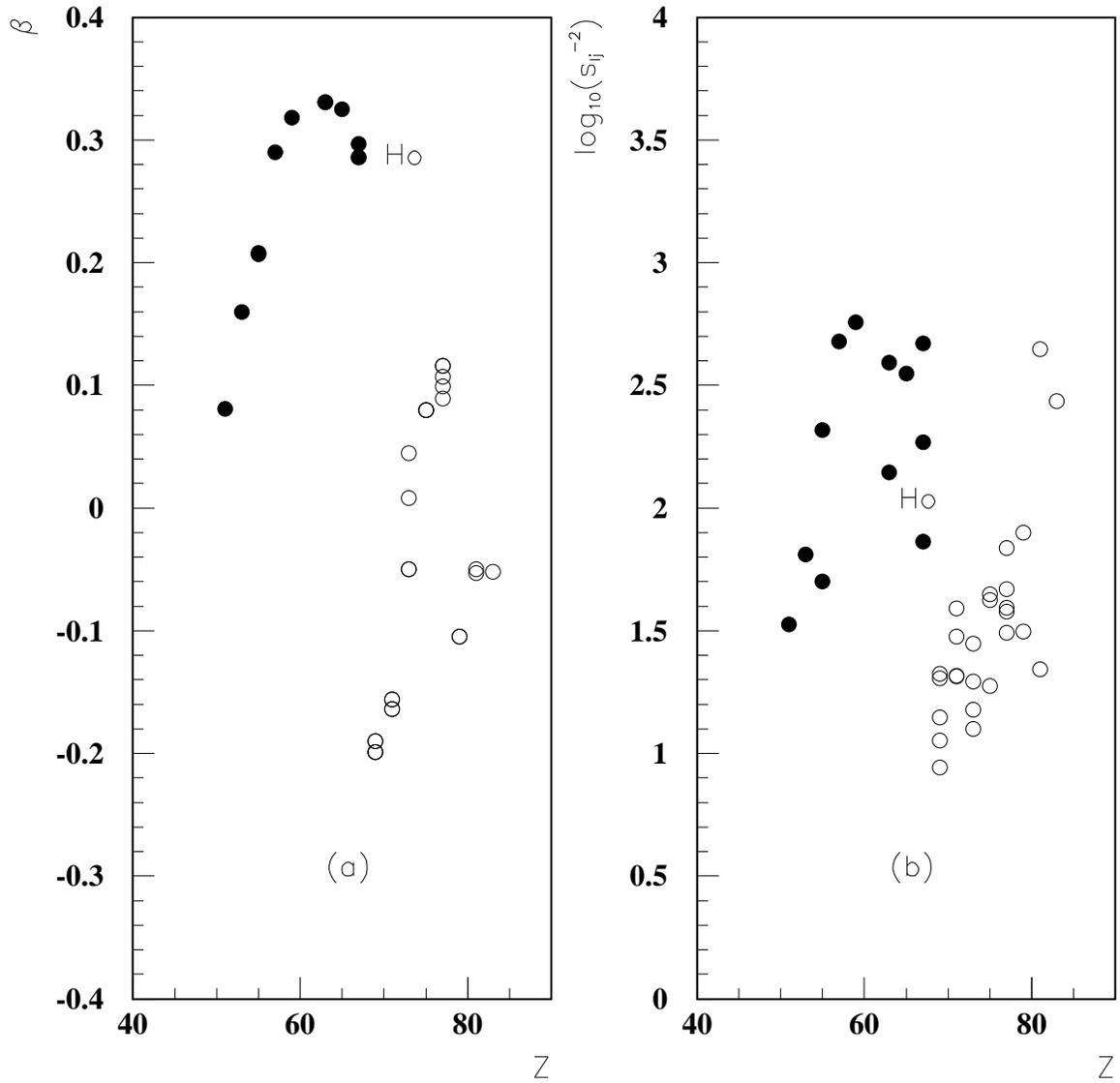}
\vspace{1cm} \caption
{
(a) Dependence of the quadrupole deformation parameter $\beta$,
as given in Table 1, upon the charge number Z.
Filled symbols correspond to the first twelve cases of
Table 1. Notice that $\beta$ is the same in this scale for
the isotopes $Cs$ and also for the two states in $Ho$.\\
(b) The logarithm of the spectroscopic function $s^{-2}_{lj}$, defined
by Eq. (\ref{T}) and given in the last column of the Table 1,
versus the charge number Z.
}
\label{fig3}
\end{center}
\end{figure}

\end{document}